\documentclass[12pt,a4paper]{article}
\usepackage{amssymb,graphicx,graphics,epsfig}
\usepackage{lscape,graphics,amsmath}
\usepackage{latexsym,amsfonts}
\usepackage{cite}
\usepackage{textcomp}
\usepackage{float}
\begin{document}
\title{\vspace{-1cm}\bf Quantization of spinor field in the\\ Schwarzschild spacetime and spin sums\\ for solutions of the Dirac equation}

\author{
Vadim Egorov, Mikhail Smolyakov, Igor Volobuev
\\
{\small{\em Skobeltsyn Institute of Nuclear Physics, Lomonosov Moscow
State University,
}}\\
{\small{\em Moscow 119991, Russia}}}

\date{}
\maketitle
\begin{abstract}
We discuss the problem of canonical quantization of a free massive
spinor field  in the Schwarzschild spacetime. It is shown that a
consistent procedure of canonical quantization of the field can be
carried out without taking into account the internal region of the
black hole, the canonical commutation relations in the resulting
theory hold exactly and the Hamiltonian has the standard form.
Spin sums are obtained for solutions of the Dirac equation in the
Schwarzschild spacetime.
\end{abstract}

\section{Introduction}
Despite the huge amount of scientific literature devoted to field
quantization in the presence of black holes, starting from
\cite{Boulware:1974dm, Hartle:1976tp} and to this day, some
questions still remain insufficiently clarified.
In particular, in the Kruskal-Szekeres
coordinates \cite{Kruskal:1959vx, Szekeres:1960gm}, which describe
the maximal analytical expansion of the Schwarzschild spacetime,
there is the so-called ``white hole'', the presence of which leads
to some problems with the physical interpretation of the resulting
theory. Indeed, there is a well-known problem
with locality associated with the location of the white hole in
our Universe or even in a parallel world. In recent papers
\cite{tHooft:2018waj, tHooft:2019xwm} an attempt was made to solve
this problem by geometrically identifying some regions in the
Kruskal-Szekeres spacetime. However, in \cite{Hooft:2022azz} it
was shown that such an identification leads to problems with CPT
invariance, and the idea of ``quantum cloning'' of the outer
regions of black and white holes was proposed instead. The inner
regions of both holes in this approach turn out to be only
mathematical artifacts that have no direct physical interpretation
and do not play any role in the evolution of the system. The
disadvantage of the described approach, however, is the
possibility of the existence of closed timelike curves
\cite{Hooft:2022azz}. In addition, there is no translational
invariance with respect to the time coordinate in the
Kruskal-Szekeres coordinates, which is also a problem, since it
does not allow one to construct a time-conserved Hamiltonian of
the system, at least if this time coordinate is interpreted as
physical time.

Another problem is that not all questions of field quantization in
the spacetime of a black hole are considered in the literature
with due mathematical rigor. For example, in  classical paper
\cite{Boulware:1974dm} the fulfillment of the commutation
relations is not checked, which is a necessary step in carrying
out the canonical quantization procedure discussed in the paper.
Moreover, to carry out a correct quantization procedure, it is
necessary to know the properties of solutions to the equations of
motion of the fields under consideration and, in particular, to
have a convenient complete system of eigenfunctions corresponding
to different quantum states, in which the quantum field will be
expanded. These issues, including the availability of
a complete set of eigenfunctions, were not discussed in detail either in \cite{Boulware:1974dm} or in
subsequent papers on this topic. Moreover, different authors make
mutually exclusive statements regarding the properties of the
spectrum of states within the framework of relativistic quantum
mechanics, even in the simplest case of a scalar field in the
gravitational field of a Schwarzschild black hole (see, for
example, \cite{Deruelle:1974zy, Zecca:2009zz}, which contradict
each other whether the spectrum of states with energies less than
the mass of the field is discrete or continuous). As far as we
know, besides \cite{Zecca:2009zz}, the only publication in which
the properties of the spectrum of the radial equation for a scalar
field are correctly described from the physical point of view is
\cite{Barranco:2011eyw}.

It should be noted that there are already papers
(see, for example, recent papers
\cite{Akhmedov:2020ryq,Anempodistov:2020oki,Bazarov:2021rrb}), in
which the quantum scalar field is considered only outside the
event horizon of the Schwarzschild black hole. Nevertheless,
despite the fact that, in some rigorous methods for constructing
quantum theory in the gravitational field of a black hole, it is
not necessary to take into account the internal regions of black
and white holes, the question remains whether it is possible to
construct a consistent quantum field theory in the Schwarzschild
spacetime only above the horizon. In paper \cite{Egorov:2022hgg},
it was rigorously shown in the case of real
scalar field that this can indeed be done even within the
framework of the standard canonical quantization. Namely, it was demonstrated that in the
corresponding quantum theory the canonical commutation relations
are satisfied exactly and the Hamiltonian has the standard form
without any pathologies. Based on this result, in the present
paper we consider the canonical quantization of a spinor field in
the Schwarzschild spacetime. It will be shown below that in this
case, too, a consistent quantum field theory can be constructed. A
particular attention will be paid to finding spin sums for
solutions of the Dirac equation, which in the case of a curved
spacetime is inseparable from the calculation of integrals of
fermion radial functions. It should be noted that equations for
the spinor field in this gravitational background and their
solutions were also studied in papers
\cite{Zecca:2007,Neznamov:2010,Neznamov:2011,Neznamov:2018zen,Neznamov:2020}.

\section{Equations of motion}
The standard metric of the Schwarzschild spacetime in
Schwarzschild coordinates looks like
\begin{equation}\label{metric_Sch}
{\rm d}s^2  = \left( {1 - \frac{{r_0 }}{r}} \right){\rm d}t^2  -
\frac{{{\rm d}r^2 }}{{1 - {\textstyle{{r_0 } \over r}}}} - r^2
\left( {{\rm d}\theta ^2  + \sin ^2 \theta \,{\rm d}\varphi ^2 }
\right),
\end{equation}
where $r_0 = 2M$ is the Schwarzschild radius and $M$ is the black
hole mass. We restrict ourselves to the region $r > r_0$ and
consider a massive spinor field.

For a remote observer, the wave functions of physical quantum states of this field above the horizon are elements of the
Hilbert space of spinor functions with finite norm, i.e., of the
field configurations $\psi(t,r,\theta, \varphi),$ which satisfy
the condition
\begin{equation} \label{norm_psi}
\int\limits_{r>r_{0}} {\frac{{r^2 \sin \theta \, {\rm d}r \, {\rm
d}\theta \, {\rm d} \varphi}}{{\sqrt {1 - {\textstyle{{r_0 } \over
r}}}
    }} \, \psi^\dagger(t,r,\theta, \varphi) \, \psi(t,r,\theta, \varphi)}  < \infty,
\end{equation}
where integration goes over a hypersurface $t={\rm const}$ in the
Schwarzschild spacetime with respect to the volume element defined
by the metric induced on this hypersurface from metric
(\ref{metric_Sch}). We emphasize that the physical states of a
quantum system are always normalizable (this point is discussed in
detail, for example, in textbook \cite{Schweber}, Chapter 1).
Since  normalization integral (\ref{norm_psi}) should be
convergent for the wave functions of the spinor particle physical
quantum states, they tend to zero at the event horizon and at
spatial infinity fast enough.

The action of the spinor field in arbitrary curvilinear
coordinates has the form
\begin{equation} \label{fermion_action}
    S = \int {L\,{\rm d}^4 x}  = \int {\sqrt { - g} \left[ {\frac{{\rm i}}{2}\left( {\bar \psi  \gamma ^{(\nu )} e_{(\nu )}^\mu  \nabla _\mu  \psi  - \nabla _\mu  \bar \psi  \,e_{(\nu )}^\mu  \gamma ^{(\nu )} \psi } \right) - m\bar \psi  \psi } \right]{\rm d}^4 x} ,
\end{equation}
where $e_{(\nu )}^\mu$ is the tetrad, $\nabla _\mu $ is the covariant derivative,
\begin{equation}
    \nabla _\mu  \psi  = \left( {\partial _\mu  \psi  + \omega _\mu  \psi } \right),\qquad \nabla _\mu  \bar \psi   = \left( {\partial _\mu  \bar \psi   - \bar \psi  \omega _\mu  } \right),
\end{equation}
\begin{equation}
    \omega _\mu   = \frac{1}{8}\omega _{(\nu )(\rho )\mu } \left[ {\gamma ^{(\nu )} ,\gamma ^{(\rho )} } \right],
\end{equation}
$\omega _{(\nu )(\rho )\mu }$ is the spin connection field,
\begin{equation}
    \omega _{(\nu )(\rho )\mu }  = g_{\sigma \tau } e_{(\nu )}^\tau  \left( {\partial _\mu  e_{(\rho )}^\sigma   + \Gamma _{\mu \lambda }^\sigma  e_{(\rho )}^\lambda  } \right),
\end{equation}
$\Gamma _{\mu \lambda }^\sigma$ are the Christoffel symbols, and
\begin{equation}
    e_{(\rho )}^\mu  \,e_{(\sigma )}^\nu  \left\{ {\gamma ^{(\rho )} ,\gamma ^{(\sigma )} } \right\} = 2g^{\mu \nu } \quad \Leftrightarrow \quad \left\{ {\gamma ^{(\mu )} ,\gamma ^{(\nu )} } \right\} = 2\eta ^{(\mu )(\nu )} .
\end{equation}

Considering action (\ref{fermion_action}) for the
physical field configurations satisfying condition
(\ref{norm_psi}) and varying it with respect to the spinor field
$\bar \psi$, we can drop the surface terms at the event horizon
and infinity and obtain the corresponding equation of motion:
\begin{equation} \label{curved_Dirac}
    {\rm i}\gamma ^{(\nu )} e_{(\nu )}^\mu  \left( {\partial _\mu   + \omega _\mu  } \right)\psi  - m\psi  = 0 .
\end{equation}
This is the Dirac equation in curvilinear coordinates.

It is a common knowledge that this Dirac equation
can be rewritten in the Hamiltonian form (see, for example,
\cite{Neznamov:2010}). We do not write out the Dirac Hamiltonian
here, because we do not need its explicit form. This Dirac
Hamiltonian is Hermitian in the Hilbert space of the physical
quantum states of the spinor particles, i.e., in the space of the
spinor functions satisfying condition (\ref{norm_psi}), which can
be easily checked by taking the matrix element of the Dirac
Hamiltonian between two normalizable states and performing
integration by parts in the terms containing derivatives (the
absence of the surface terms is guaranteed by
tending to zero of the wave functions at the event horizon and at
spatial infinity). However, it is necessary to
note that the eigenfunctions of the Dirac Hamiltonian need not to
belong to the Hilbert space of normalizable spinor functions. They
may  have infinite norm, i.e., they can lie in the so-called
rigged Hilbert space, that is, they are generalized functions
(see, for example, \cite{rigged_Hilbert} and references therein).
This situation is similar, for example, to the description of a
free particle in non-relativistic quantum mechanics: the wave
functions of the physical states are normalizable, whereas  the
eigenfunctions of the free Hamiltonian, which can be chosen as
plane waves, i.e. the eigenfunctions of the momentum operator, are
not normalizable. They are not physical quantum states, the latter
being realized by normalizable wave packets. Nevertheless, the
plane waves are very convenient for calculations, because their
scalar products are very simple, they make up a complete system of eigenfunctions and we can
expand normalizable physical states in this system.

Thus, we look for classical stationary solutions
to equation (\ref{curved_Dirac}), which turn out to be the
eigenfunctions of the corresponding Dirac Hamiltonian with
infinite norm. Since the Dirac Hamiltonian is Hermitian, the
system of these solutions to Dirac equation (\ref{curved_Dirac})
will also make up a complete system of eigenfunctions. Because
metric (\ref{metric_Sch}) is spherically symmetric,
the classical solutions are characterized by the total angular
momentum and include the spherical spinors as their angular part.
However, the standard choice of the tetrad in Schwarzschild
coordinates, the so-called Schwinger gauge, turns out to be very
inconvenient for working with the spherical spinors
\cite{Boulware:1975pe,Neznamov:2018zen}. For this reason, we first
consider equation (\ref{curved_Dirac}) in the isotropic
coordinates, which allows us to use the standard technique for
working with spherical spinors. After the corresponding
calculations are completed, we will pass to Schwarzschild
coordinates. The transition from the isotropic coordinates
$\left\{ {t,x,y,z } \right\}$  to Schwarzschild coordinates
$\left\{ {t,r,\theta ,\varphi } \right\}$  is carried out
according to the formula
\begin{equation}
    r = R + \frac{{M^2 }}{{4R}} + M = R + \frac{{r_0^2 }}{{16R}} +
    \frac{{r_0}}{2}, \quad R = \sqrt{x^2  + y^2  + z^2}.
\end{equation}

 The components of the metric in the isotropic coordinates are
\begin{equation}
\begin{gathered}
  g_{00}  = \frac{{\left( {1 - {\textstyle{{r_0 } \over {4R}}}} \right)^2 }}{{\left( {1 + {\textstyle{{r_0 } \over {4R}}}} \right)^2 }},\quad g_{kk}  =  - \left( {1 + \frac{{r_0 }}{{4R}}} \right)^4 ,\\
  g_{\mu \nu }  = 0,\quad \mu  \ne \nu .
\end{gathered}
\end{equation}
Here and below, the Latin coordinate index $k$ corresponds to the spatial coordinates, i.e.\ runs through the values 1, 2, 3. We define the tetrad as follows:
\begin{equation}
\begin{gathered}
 e_{(0)}^0  = \frac{{1 + {\textstyle{{r_0 } \over {4R}}}}}{{1 - {\textstyle{{r_0 } \over {4R}}}}},\quad e_{(k)}^k  = \left( {1 + \frac{{r_0 }}{{4R}}} \right)^{ - 2} , \\
 e_{(\nu )}^\mu   = 0,\quad \mu  \ne \nu .
\end{gathered}
\end{equation}
The Christoffel symbols in the isotropic coordinates are:
\begin{equation}
    \Gamma _{\mu \nu }^0  = \frac{{r_0 }}{{2R^3 }}\frac{1}{{\left( {1 - {\textstyle{{r_0 } \over {4R}}}} \right) \left( {1 + {\textstyle{{r_0 } \over {4R}}}} \right)}}\left( {\begin{array}{*{20}c}
   0 & x & y & z  \\
   x & 0 & 0 & 0  \\
   y & 0 & 0 & 0  \\
   z & 0 & 0 & 0  \\
\end{array}} \right),
\end{equation}
\begin{equation}
    \Gamma _{\mu \nu }^1  = \frac{{r_0 }}{{2R^3 }}\frac{1}{{1 + {\textstyle{{r_0 } \over {4R}}}}}\left( {\begin{array}{*{20}c}
   {\frac{{1 - {\textstyle{{r_0 } \over {4R}}}}}{{\left( {1 + {\textstyle{{r_0 } \over {4R}}}} \right)^6 }}x} & 0 & 0 & 0  \\
   0 & { - x} & { - y} & { - z}  \\
   0 & { - y} & x & 0  \\
   0 & { - z} & 0 & x  \\
\end{array}} \right),
\end{equation}
\begin{equation}
    \Gamma _{\mu \nu }^2  = \frac{{r_0 }}{{2R^3 }}\frac{1}{{1 + {\textstyle{{r_0 } \over {4R}}}}}\left( {\begin{array}{*{20}c}
   {\frac{{1 - {\textstyle{{r_0 } \over {4R}}}}}{{\left( {1 + {\textstyle{{r_0 } \over {4R}}}} \right)^6 }}y} & 0 & 0 & 0  \\
   0 & y & { - x} & 0  \\
   0 & { - x} & { - y} & { - z}  \\
   0 & 0 & { - z} & y  \\
\end{array}} \right),
\end{equation}
\begin{equation}
    \Gamma _{\mu \nu }^3  = \frac{{r_0 }}{{2R^3 }}\frac{1}{{1 + {\textstyle{{r_0 } \over {4R}}}}}\left( {\begin{array}{*{20}c}
   {\frac{{1 - {\textstyle{{r_0 } \over {4R}}}}}{{\left( {1 + {\textstyle{{r_0 } \over {4R}}}} \right)^6 }}z} & 0 & 0 & 0  \\
   0 & z & 0 & { - x}  \\
   0 & 0 & z & { - y}  \\
   0 & { - x} & { - y} & { - z}  \\
\end{array}} \right).
\end{equation}
Substituting all this into the spin connection, we get
\begin{equation}
\omega _0   = \frac{{r_0 }}{{8R^3 }}\left( {1 + \frac{{r_0
}}{{4R}}} \right)^{ - 4} { \left[ {\gamma ^{(0)} ,\vec x\vec
\gamma } \right]}, \quad \omega _k   = \frac{{r_0 }}{{8R^3
}}\left( {1 + \frac{{r_0 }}{{4R}}} \right)^{ - 1} { \left[ {
\gamma^{(k)} ,\vec x\vec \gamma } \right]},
\end{equation}
where  $\vec \gamma  \equiv \left( {\gamma ^{(1)} ,\gamma ^{(2)}
,\gamma ^{(3)} } \right)$,   $\vec x\vec \gamma  \equiv x^1 \gamma
^{(1)}  + x^2 \gamma ^{(2)}  + x^3 \gamma ^{(3)}  = x\gamma ^{(1)}
+ y\gamma ^{(2)}  + z\gamma ^{(3)} $.

Turning to  Dirac equation (\ref{curved_Dirac}), we have
\begin{equation}
\begin{split}
 {\rm i}\frac{{1 + {\textstyle{{r_0 } \over {4R}}}}}{{1 - {\textstyle{{r_0 } \over {4R}}}}}\gamma ^{(0)} \partial _0 \psi  + {\rm i}\left( {1 + \frac{{r_0 }}{{4R}}} \right)^{ - 2} \left( {\gamma ^{(1)} \partial _1  + \gamma ^{(2)} \partial _2  + \gamma ^{(3)} \partial _3 } \right)\psi  +  \hphantom{= 0 .} \\
\qquad  + \, {\rm i}\frac{{r_0 }}{{4R^3 }}\left( {1 + \frac{{r_0 }}{{4R}}} \right)^{ - 3} \left[ {\left( {1 - \frac{{r_0 }}{{4R}}} \right)^{ - 1}  - 2} \right]\left( {\vec x\vec \gamma } \right)\psi  - m\psi  = 0.
\end{split}
\end{equation}
We represent this equation for the bispinor $\psi$ as two
equations for spinors $\xi$ and $\chi$,
\begin{equation}
    \psi  = \left( {\begin{array}{*{20}c}
   \xi   \\
   \chi   \\
\end{array}} \right).
\end{equation}
To do this, we need explicit expressions for the
gamma matrices in terms of the Pauli matrices. In accordance with
monograph \cite{AB}, we choose the standard representation of
gamma matrices (the Pauli-Dirac representation):
\begin{equation}
    \gamma ^{(0)}  = \left( {\begin{array}{*{20}c}
   1 & 0  \\
   0 & { - 1}  \\
\end{array}} \right),\qquad \gamma ^{(k)}  = \left( {\begin{array}{*{20}c}
   0 & {\sigma ^{(k)} }  \\
   { - \sigma ^{(k)} } & 0  \\
\end{array}} \right),
\end{equation}
where $\sigma ^{(k)}$ are the Pauli matrices. In this way, we get
the system of equations
\begin{equation} \label{EoM_xi_chi}
\begin{split}
 &{\rm i}\frac{{1 + {\textstyle{{r_0 } \over {4R}}}}}{{1 - {\textstyle{{r_0 } \over {4R}}}}}\frac{{\partial \xi }}{{\partial t}} - \left( {1 + \frac{{r_0 }}{{4R}}} \right)^{ - 2} \left( {\vec \sigma \vec p} \right)\chi \, + \\
 &+ {\rm i}\frac{{r_0 }}{{4R^3 }}\left( {1 + \frac{{r_0 }}{{4R}}} \right)^{ - 3} \left[ {\left( {1 - \frac{{r_0 }}{{4R}}} \right)^{ - 1}  - 2} \right]\left( {\vec x\vec \sigma } \right)\chi  - m\xi  = 0, \\
 &{\rm i}\frac{{1 + {\textstyle{{r_0 } \over {4R}}}}}{{1 - {\textstyle{{r_0 } \over {4R}}}}}\frac{{\partial \chi }}{{\partial t}} - \left( {1 + \frac{{r_0 }}{{4R}}} \right)^{ - 2} \left( {\vec \sigma \vec p} \right)\xi \, + \\
 &+ {\rm i}\frac{{r_0 }}{{4R^3 }}\left( {1 + \frac{{r_0 }}{{4R}}} \right)^{ - 3} \left[ {\left( {1 - \frac{{r_0 }}{{4R}}} \right)^{ - 1}  - 2} \right]\left( {\vec x\vec \sigma } \right)\xi  + m\chi  = 0,
\end{split}
\end{equation}
where
\begin{equation}
    \left( {\vec \sigma \vec p} \right) =  - {\rm i}\left( {\vec \sigma \nabla } \right) =  - {\rm i}\left( {\sigma ^{(1)} \frac{\partial }{{\partial x}} + \sigma ^{(2)} \frac{\partial }{{\partial y}} + \sigma ^{(3)} \frac{\partial }{{\partial z}}} \right) =  - {\rm i}\left( {\sigma ^{(1)} \partial _1  + \sigma ^{(2)} \partial _2  + \sigma ^{(3)} \partial _3 } \right),
\end{equation}
\begin{equation}
    \left( {\vec x\vec \sigma } \right) \equiv x^1 \sigma ^{(1)}  + x^2 \sigma ^{(2)}  + x^3 \sigma ^{(3)}  = x\sigma ^{(1)}  + y\sigma ^{(2)}  + z\sigma ^{(3)} .
\end{equation}

\section{Spectrum of stationary states}
Let us pass from the Cartesian coordinates $\left\{ {x,y,z}
\right\}$ to the spherical coordinates $\left\{ {R,\theta ,\varphi
} \right\}$, where $R = \sqrt {x^2  + y^2  + z^2 } $ is still the
same.

Taking into account the spherical symmetry of equations
(\ref{EoM_xi_chi}), we  look for  stationary solutions to them in
the form
\begin{equation}
    \psi _{Ejlm_{\rm t} } \left( {t,R,\theta ,\varphi } \right) = \left( {\begin{array}{*{20}c}
   {F_{jl} \left( {E,R} \right)\Omega _{jlm_{\rm t} } \left( {\theta ,\varphi } \right)}  \\
   {{\rm i}G_{jl'} \left( {E,R} \right)\Omega _{jl'm_{\rm t} } \left( {\theta ,\varphi } \right)}  \\
\end{array}} \right){\rm e}^{ - {\rm i}Et} ,
\end{equation}
where $j$ is the total angular momentum, $m_{\rm t}$ is its
projection; $l = j \pm \frac{1}{2}$, $l' = j \mp \frac{1}{2}$ are
the orbital angular momenta;
\begin{equation}
    \Omega _{jlm_{\rm t} } \left( {\theta ,\varphi } \right) = \left( {\begin{array}{*{20}c}
   {C_{l,m_{\rm t}  - {\textstyle{1 \over 2}},{\textstyle{1 \over 2}},{\textstyle{1 \over 2}}}^{jm_{\rm t} } Y_l^{m_{\rm t}  - {\textstyle{1 \over 2}}} \left( {\theta ,\varphi } \right)}  \\
   {C_{l,m_{\rm t}  + {\textstyle{1 \over 2}},{\textstyle{1 \over 2}}, - {\textstyle{1 \over 2}}}^{jm_{\rm t} } Y_l^{m_{\rm t}  + {\textstyle{1 \over 2}}} \left( {\theta ,\varphi } \right)}  \\
\end{array}} \right)
\end{equation}
are the spherical spinors, where the Clebsch-Gordan coefficients
$C_{l,m_{\rm t}  \mp {\textstyle{1 \over 2}},{\textstyle{1 \over
2}}, \pm {\textstyle{1 \over 2}}}^{jm_{\rm t} }$ can be found, for
example, in  monograph \cite{AB}. The spherical spinors $\Omega
_{jlm_{\rm t} } \left( {\theta ,\varphi } \right)$ form a complete
orthogonal system of functions in the space of complex spinors on
the sphere $S^2$:
\begin{equation} \label{norm_Omega}
    \int {\Omega _{jlm_{\rm t} }^ \dagger  \left( {\theta ,\varphi } \right)\Omega _{\tilde j \tilde l \tilde m_{\rm t} } \left( {\theta ,\varphi } \right)\sin \theta \,{\rm d}\theta \,{\rm d}\varphi }  = \delta _{j \tilde j} \,\delta _{l \tilde l} \,\delta _{m_{\rm t} \tilde m_{\rm t} }
\end{equation}
(orthogonality),
\begin{equation} \label{completeness_Omega}
    \sum\limits_{j = {\textstyle{1 \over 2}}}^\infty  {\sum\limits_{m_{\rm t}  =  - j}^j {\sum\limits_{l = j \pm {\textstyle{1 \over 2}}} {\Omega _{jlm_{\rm t} } \left( {\theta ,\varphi } \right)\Omega _{jlm_{\rm t} }^ \dagger  \left( {\theta ',\varphi '} \right)} } }  = \left( {\begin{array}{*{20}c}
   1 & 0  \\
   0 & 1  \\
\end{array}} \right)\delta \left( {\cos \theta  - \cos \theta '} \right)\delta \left( {\varphi  - \varphi '} \right)
\end{equation}
(completeness).

Given that \cite{AB}
\begin{equation}
\begin{split}
    &\left( {\vec \sigma \vec p} \right)\left[ {F_{jl} \left( {E,R} \right)\Omega _{jlm_{\rm t} } \left( {\theta ,\varphi } \right)} \right] =  \\
  &=  - {\rm i}\frac{{{\rm d}F_{jl} \left( {E,R} \right)}}{{{\rm d}R}}\frac{1}{R}\left( {\vec \sigma \vec x} \right)\,\Omega _{jlm_{\rm t} } \left( {\theta ,\varphi } \right) + F_{jl} \left( {E,R} \right)\left( {\vec \sigma \vec p} \right)\Omega _{jlm_{\rm t} } \left( {\theta ,\varphi } \right),
\end{split}
\end{equation}
\begin{equation}
\begin{split}
 &\left( {\vec \sigma \vec p} \right)\left[ {{\rm i}G_{jl'} \left( {E,R} \right)\Omega _{jl'm_{\rm t} } \left( {\theta ,\varphi } \right)} \right] =  \\
 &= \frac{{{\rm d}G_{jl'} \left( {E,R} \right)}}{{{\rm d}R}}\frac{1}{R}\left( {\vec \sigma \vec x} \right)\Omega _{jl'm_{\rm t} } \left( {\theta ,\varphi } \right) + {\rm i}G_{jl'} \left( {E,R} \right)\left( {\vec \sigma \vec p} \right)\Omega _{jl'm_{\rm t} } \left( {\theta ,\varphi } \right),
\end{split}
\end{equation}
\begin{equation}
    \frac{1}{R}\left( {\vec \sigma \vec x} \right)\Omega _{jlm_{\rm t} } \left( {\theta ,\varphi } \right) =  - \Omega _{jl'm_{\rm t} } \left( {\theta ,\varphi } \right),\qquad l + l' = 2j,
\end{equation}
\begin{equation}
    \left( {\vec \sigma \vec p} \right)\Omega _{jlm_{\rm t} } \left( {\theta ,\varphi } \right) = {\rm i}\frac{{1 + \kappa }}{R}\Omega _{jl'm_{\rm t} } \left( {\theta ,\varphi } \right),
\end{equation}
\begin{equation}
    \left( {\vec \sigma \vec p} \right)\Omega _{jl'm_{\rm t} } \left( {\theta ,\varphi } \right) = {\rm i}\frac{{1 - \kappa }}{R}\Omega _{jlm_{\rm t} } \left( {\theta ,\varphi } \right),
\end{equation}
where $\kappa  = l\left( {l + 1} \right) - j\left( {j + 1} \right) - \frac{1}{4}$, from equations (\ref{EoM_xi_chi}) we get
\begin{equation} \label{Radial_F}
\begin{split}
    &\left( {1 + \frac{{r_0 }}{{4R}}} \right)^{ - 2} \left( {\frac{{{\rm d}F_{jl} \left( {E,R} \right)}}{{{\rm d}R}} + \frac{{1 + \kappa }}{R}F_{jl} \left( {E,R} \right)} \right) +  \\
  &+ \frac{{r_0 }}{{4R^2 }}\left( {1 + \frac{{r_0 }}{{4R}}} \right)^{ - 3} \left[ {\left( {1 - \frac{{r_0 }}{{4R}}} \right)^{ - 1}  - 2} \right]F_{jl} \left( {E,R} \right) - \left( {\frac{{1 + {\textstyle{{r_0 } \over {4R}}}}}{{1 - {\textstyle{{r_0 } \over {4R}}}}}E + m} \right)G_{jl'} \left( {E,R} \right) = 0,
\end{split}
\end{equation}
\begin{equation} \label{Radial_G}
\begin{split}
    &\left( {1 + \frac{{r_0 }}{{4R}}} \right)^{ - 2} \left( {\frac{{{\rm d}G_{jl'} \left( {E,R} \right)}}{{{\rm d}R}} + \frac{{1 - \kappa }}{R}G_{jl'} \left( {E,R} \right)} \right) +  \\
  &+ \frac{{r_0 }}{{4R^2 }}\left( {1 + \frac{{r_0 }}{{4R}}} \right)^{ - 3} \left[ {\left( {1 - \frac{{r_0 }}{{4R}}} \right)^{ - 1}  - 2} \right]G_{jl'} \left( {E,R} \right) + \left( {\frac{{1 + {\textstyle{{r_0 } \over {4R}}}}}{{1 - {\textstyle{{r_0 } \over {4R}}}}}E - m} \right)F_{jl} \left( {E,R} \right) = 0.
\end{split}
\end{equation}

Let us discuss the asymptotic behavior of these equations for $R \to \infty $. They take the form
\begin{equation}
    \frac{{{\rm d}F_{jl} \left( {E,R} \right)}}{{{\rm d}R}} + \frac{{1 + \kappa }}{R}F_{jl} \left( {E,R} \right) - \left( {E + m + \frac{{r_0 }}{{2R}}E} \right)G_{jl'} \left( {E,R} \right) = 0,
\end{equation}
\begin{equation}
    \frac{{{\rm d}G_{jl'} \left( {E,R} \right)}}{{{\rm d}R}} + \frac{{1 - \kappa }}{R}G_{jl'} \left( {E,R} \right) + \left( {E - m + \frac{{r_0 }}{{2R}}E} \right)F_{jl} \left( {E,R} \right) =
    0,
\end{equation}
which can be rewritten as
\begin{equation}
    \frac{{{\rm d}\left(R F_{jl} \left( {E,R}\right) \right)}}{{{\rm d}R}} + \frac{\kappa }{R}\left(R F_{jl} \left( {E,R}\right) \right) - \left( {E + m + \frac{{r_0 }}{{2R}}E} \right)\left(R G_{jl'} \left( {E,R}\right) \right) = 0,
\end{equation}
\begin{equation}
    \frac{{{\rm d}\left(R G_{jl'} \left( {E,R}\right) \right)}}{{{\rm d}R}} - \frac{\kappa }{R}\left(R G_{jl'} \left( {E,R}\right) \right) + \left( {E - m + \frac{{r_0 }}{{2R}}E} \right)\left(R F_{jl} \left( {E,R}\right) \right) = 0.
\end{equation}
The last equations coincide with those for an electron in the
Coulomb potential of an atomic nucleus, presented in  monograph
\cite{AB}, up to the redefinition $F_{jl} \left( {E,R} \right) \to
g\left( R \right)$,  $G_{jl'} \left( {E,R} \right) \to f\left( R
\right)$. Comparing the signs in the last terms, we notice that,
as in the case of the Coulomb field, the resulting potential is
attractive:
\begin{equation}
    V\left( R \right) =  - \frac{{r_0 }}{{2R}}E =  - \frac{M}{R}E,
\end{equation}
where $M = {{r_0 } \mathord{\left/ {\vphantom {{r_0 } 2}} \right.
\kern-\nulldelimiterspace} 2}$ is the black hole mass.  This
potential reproduces the Newtonian one with the only  difference
that it is proportional to the total energy $E$ of the particle
instead of its mass $m$. This happens due to the relativistic
nature of the equation, since in general relativity the source of
gravity is precisely the energy-momentum.

Let us return to the Schwarzschild coordinates in full radial
equations (\ref{Radial_F}) and (\ref{Radial_G}) for the spinor field
outside the black hole horizon:
\begin{equation} \label{radial_F}
    \sqrt {1 - \frac{{r_0 }}{r}} \frac{{{\rm d}F_{jl} }}{{{\rm d}r}} + \left( {\frac{{\sqrt {1 - {\textstyle{{r_0 } \over r}}}  + \kappa }}{r} + \frac{{r_0 }}{{4r^2 \sqrt {1 - {\textstyle{{r_0 } \over r}}} }}} \right)F_{jl}  - \left( {\frac{E}{{\sqrt {1 - {\textstyle{{r_0 } \over r}}} }} + m} \right)G_{jl'}  = 0,
\end{equation}
\begin{equation} \label{radial_G}
    \sqrt {1 - \frac{{r_0 }}{r}} \frac{{{\rm d}G_{jl'} }}{{{\rm d}r}} + \left( {\frac{{\sqrt {1 - {\textstyle{{r_0 } \over r}}}  - \kappa }}{r} + \frac{{r_0 }}{{4r^2 \sqrt {1 - {\textstyle{{r_0 } \over r}}} }}} \right)G_{jl'}  + \left( {\frac{E}{{\sqrt {1 - {\textstyle{{r_0 } \over r}}} }} - m} \right)F_{jl}  = 0,
\end{equation}
where $r \in \left( {r_0 ,\infty } \right)$. These equations can be used to show that the orthogonality condition for the solutions has the form
\begin{equation} \label{norm_FG}
   \int\limits_{r_0}^\infty {\frac{{r^2 {\rm d}r}}{{\sqrt {1 - {\textstyle{{r_0 } \over r}}} }}\left( {F_{jl} \left( {E,r} \right)\,F_{jl} \left( {E',r} \right) + G_{jl'} \left( {E,r} \right)\,G_{jl'} \left( {E',r} \right)} \right)}  = \delta \left( {E - E'} \right).
\end{equation}
Obviously, the integral in this formula is the
radial part of the integral in normalization condition
(\ref{norm_psi}). Condition (\ref{norm_FG}) implies that the
second-order equations for the functions $F_{jl}$ and $G_{jl'}$
cannot be reduced to equations in the Sturm-Liouville form,
otherwise it would be possible to obtain orthonormality conditions
for the functions $F_{jl}$ and $G_{jl'}$ separately. Without loss
of generality, the functions $F_{jl}$ and $G_{jl'}$ can be
considered real.

Let us make the substitution
\begin{equation}
    F_{jl}  = \frac{{f_{jl} }}{{r\left( {1 - {\textstyle{{r_0 } \over r}}} \right)^{{\textstyle{1 \over 4}}} }},\qquad G_{jl'}  = \frac{{g_{jl'} }}{{r\left( {1 - {\textstyle{{r_0 } \over r}}} \right)^{{\textstyle{1 \over 4}}} }}.
\end{equation}
After such substitution, equations (\ref{radial_F}) and
(\ref{radial_G}) take the simpler form
\begin{equation} \label{radial_f}
    \sqrt {1 - \frac{{r_0 }}{r}} \frac{{{\rm d}f_{jl} }}{{{\rm d}r}} + \frac{\kappa }{r}f_{jl}  - \left( {\frac{E}{{\sqrt {1 - {\textstyle{{r_0 } \over r}}} }} + m} \right)g_{jl'}  = 0,
\end{equation}
\begin{equation} \label{radial_g}
    \sqrt {1 - \frac{{r_0 }}{r}} \frac{{{\rm d}g_{jl'} }}{{{\rm d}r}} - \frac{\kappa }{r}g_{jl'}  + \left( {\frac{E}{{\sqrt {1 - {\textstyle{{r_0 } \over r}}} }} - m} \right)f_{jl}  = 0
\end{equation}
(perhaps this is the simplest form that the original equations
(\ref{radial_F}) and (\ref{radial_G}) can be reduced to). The
normalization conditions for these functions can be written as
\begin{equation} \label{norm_fg}
   \int\limits_{r_0}^\infty {\frac{{{\rm d}r}}{{1 - {\textstyle{{r_0 } \over r}}}}\left( {f_{jl} \left( {E,r} \right)f_{jl} \left( {E',r} \right) + g_{jl'} \left( {E,r} \right)g_{jl'} \left( {E',r} \right)} \right)}  = \delta \left( {E - E'} \right).
\end{equation}
Equations (\ref{radial_f}) and (\ref{radial_g}) can be written in the matrix form
\begin{equation}
    \left( { - {\rm i}\sigma ^{(2)} \left( {1 - \frac{{r_0 }}{r}} \right)\frac{{\rm d}}{{{\rm d}r}} + \sigma ^{(1)} \frac{{\kappa \sqrt {1 - {\textstyle{{r_0 } \over r}}} }}{r} + \sigma ^{(3)} \sqrt {1 - \frac{{r_0 }}{r}} m} \right)\left( {\begin{array}{*{20}c}
   {f_{jl} }  \\
   {g_{jl'} }  \\
\end{array}} \right) = E\left( {\begin{array}{*{20}c}
   {f_{jl} }  \\
   {g_{jl'} }  \\
\end{array}} \right).
\end{equation}
Thus, we obtain the eigenvalue problem for the operator
\begin{equation}
    H_r \left( \kappa  \right) =  - {\rm i}\sigma ^{(2)} \left( {1 - \frac{{r_0 }}{r}} \right)\frac{{\rm d}}{{{\rm d}r}} + \sigma ^{(1)} \frac{{\kappa \sqrt {1 - {\textstyle{{r_0 } \over r}}} }}{r} + \sigma ^{(3)} \sqrt {1 - \frac{{r_0 }}{r}} m
\end{equation}
in the space of real spinor functions of $r$,   $\xi \left( r
\right) = \left( {\begin{array}{*{20}c}   {\xi _1 \left( r
\right)}  \\   {\xi _2 \left( r \right)}  \\ \end{array}}
\right),$ with the scalar product
\begin{equation} \label{scal_prod}
    \int\limits_{r_0}^\infty {\frac{{{\rm d}r}}{{1 - {\textstyle{{r_0 } \over r}}}}\eta^{\rm T}  \left( r \right)\xi \left( r \right)}  = \int\limits_{r_0}^\infty {\frac{{{\rm d}r}}{{1 - {\textstyle{{r_0 } \over r}}}}\left( {\eta _1 \left( r \right)\xi _1 \left( r \right) + \eta _2 \left( r \right)\xi _2 \left( r \right)} \right)} .
\end{equation}

It is easy to check that the operator $H_r \left( \kappa  \right)$
is Hermitian in the Hilbert space of the real spinor functions
with finite norm defined by this scalar product.
Therefore, its eigenfunctions form a complete orthogonal system in
the space of real spinors with scalar product (\ref{scal_prod}).

Note that if $\xi  = \left( {\begin{array}{*{20}c}   {f_{jl} }  \\
{g_{jl'} }  \\ \end{array}} \right)$ is the eigenfunction of the
operator $H_r \left( \kappa  \right)$ with an eigenvalue $E$,
i.e.\ $H_r \left( \kappa  \right)\xi  = E\xi$, then $\sigma ^{(1)}
\xi  = \left( {\begin{array}{*{20}c}   {g_{jl'} }  \\   {f_{jl} }
\\ \end{array}} \right)$ is the eigenfunction of the operator
\begin{equation}
    H_r \left( { - \kappa } \right) =  - \sigma ^{(1)} H_r \left( \kappa  \right)\sigma ^{(1)}  =  - {\rm i}\sigma ^{(2)} \left( {1 - \frac{{r_0 }}{r}} \right)\frac{{\rm d}}{{{\rm d}r}} - \sigma ^{(1)} \frac{{\kappa \sqrt {1 - {\textstyle{{r_0 } \over r}}} }}{r} + \sigma ^{(3)} \sqrt {1 - \frac{{r_0 }}{r}} m
\end{equation}
corresponding to the eigenvalue $-E$.  Therefore, the
eigenfunctions of $H_r \left( \kappa  \right)$ corresponding to
the eigenvalues $-E$, $E > 0$, will be the functions $\left(
{\begin{array}{*{20}c}   {g_{jl} \left( {E,r} \right)}  \\
{f_{jl'} \left( {E,r} \right)}  \\ \end{array}} \right)$. With
this in mind, from orthogonality condition (\ref{norm_fg}) of the
eigenfunctions, we find that
\begin{equation}
    \int\limits_{r_0}^\infty {\frac{{{\rm d}r}}{{1 - {\textstyle{{r_0 } \over r}}}}\left( {f_{jl} \left( {E,r} \right)g_{jl} \left( {E,r} \right) + f_{jl'} \left( {E,r} \right)g_{jl'} \left( {E,r} \right)} \right)}  = 0.
\end{equation}

Passing to the second-order equations for $f_{jl}$ or $g_{jl'}$,
below we will show that the spectrum of the
operator $H_r \left( \kappa \right)$ is continuous and the
eigenvalues run through the entire set of real numbers $\mathbb{R}$, i.e.\
$- \infty  < E < \infty$.

For further analysis of equations
(\ref{radial_f}) and (\ref{radial_g}),  it is convenient to introduce
the dimensionless variables $\rho = \frac{r}{r_0}$, $\varepsilon
= r_0 E$, $\mu  = r_0 m$, that reduce the equations to the form
\begin{equation} \label{dim_less_f}
    \sqrt {1 - \frac{1}{\rho }} \frac{{{\rm d}f_{jl} }}{{{\rm d}\rho }} + \frac{\kappa }{\rho }f_{jl}  - \left( {\frac{\varepsilon }{{\sqrt {1 - {\textstyle{1 \over \rho }}} }} + \mu } \right)g_{jl'}  = 0,
\end{equation}
\begin{equation}\label{dim_less_g}
    \sqrt {1 - \frac{1}{\rho }} \frac{{{\rm d}g_{jl'} }}{{{\rm d}\rho }} - \frac{\kappa }{\rho }g_{jl'}  + \left( {\frac{\varepsilon }{{\sqrt {1 - {\textstyle{1 \over \rho }}} }} - \mu } \right)f_{jl}  = 0.
\end{equation}

First, let us consider the case $\varepsilon  > 0$. We
differentiate equation (\ref{dim_less_f})  with respect to $\rho$
and use original equations (\ref{dim_less_f}) and
(\ref{dim_less_g}). After that, the equation takes the form (in
what follows, for brevity, we omit the indices
$j$, $l$, $l'$ of the radial functions, where this does not lead
to a confusion):
\begin{equation} \label{2_order_f_rho}
    \frac{{{\rm d}^{\rm 2} f}}{{{\rm d}\rho ^2 }} + w\left( \rho  \right)\frac{{{\rm d}f}}{{{\rm d}\rho }} + c\left( \rho  \right)f = 0,
\end{equation}
where
\begin{equation}\label{singular_potential}
\begin{split}
w\left( \rho  \right) &= \frac{1}{{2\rho \left( {\rho  - 1}
\right)}} + \frac{\varepsilon }{{2\rho \left( {\rho  - 1}
\right)\left( {\varepsilon  + \mu
\sqrt {{\textstyle{{\rho  - 1} \over \rho }}} } \right)}}, \\
c\left( \rho \right) &= \frac{{\varepsilon ^2 \rho ^2 }}{{\left(
{\rho  - 1} \right)^2 }} - \frac{{\mu ^2 \rho }}{{\rho  - 1}} -
\frac{\kappa }{{\rho ^{{\textstyle{3 \over 2}}} \sqrt {\rho  - 1}
}} - \frac{{\kappa ^2 }}{{\rho \left( {\rho  - 1} \right)}} +
\frac{{\kappa \varepsilon }}{{2\rho ^{{\textstyle{3 \over 2}}}
\left( {\rho  - 1} \right)^{{\textstyle{3 \over 2}}} \left(
{\varepsilon  + \mu \sqrt {{\textstyle{{\rho  - 1} \over \rho }}}
} \right)}}.
\end{split}
\end{equation}

Let us pass to the tortoise coordinate $\rho \left( z \right) + \ln \left( {\rho \left( z \right) - 1} \right) = z$. After the transformation of the coordinate $\rho \to z$, equation (\ref{2_order_f_rho}) takes the form
\begin{equation} \label{2_order_f_z}
    \frac{{{\rm d}^{\rm 2} f}}{{{\rm d}z^2 }} + W\left( z \right)\frac{{{\rm d}f}}{{{\rm d}z}} + C\left( z \right)f = 0 ,
\end{equation}
where
\begin{equation}
\begin{split}
  W\left( z \right) &= w\left( {\rho \left( z \right)} \right)\frac{{{\rm d}\rho \left( z \right)}}{{{\rm d}z}} - \frac{{{\rm d}^{\rm 2} \rho \left( z \right)}}{{{\rm d}z^2 }}\left( {\frac{{{\rm d}\rho \left( z \right)}}{{{\rm d}z}}} \right)^{ - 1} , \\
  C\left( z \right) &= c\left( {\rho \left( z \right)} \right)\,\left( {\frac{{{\rm d}\rho \left( z \right)}}{{{\rm d}z}}} \right)^2 .
\end{split}
\end{equation}
In the coordinate $z$, orthonormality condition (\ref{norm_FG})
rewritten in terms of the functions $f_{jl}$ and $g_{jl'}$ takes
the form
\begin{equation}
\begin{split}
    &\int\limits_{ - \infty }^\infty  {\left( {f_{jl}\left( {\varepsilon ,z} \right)f_{jl}\left( {\varepsilon ',z} \right) + g_{jl'}\left( {\varepsilon ,z} \right)g_{jl'}\left( {\varepsilon ',z} \right)} \right){\rm d}z}  =  \\
  &= \int\limits_{ - \infty }^\infty  {\left( {f_{jl}\left( {\varepsilon ,z} \right),\,\,g_{jl'}\left( {\varepsilon ,z} \right)} \right) \cdot \left( {\begin{array}{*{20}c}
   {f_{jl}\left( {\varepsilon ',z} \right)}  \\
   {g_{jl'}\left( {\varepsilon ',z} \right)}  \\
\end{array}} \right){\rm d}z}  = \delta \left( {\varepsilon  - \varepsilon '} \right).
\end{split}
\end{equation}
The last step is to bring equation (\ref{2_order_f_z}) to the form of the Schr\"odinger equation. To this end, we make one more substitution
\begin{equation} \label{f_Au}
    f = A\left( z \right)\,u, \qquad A\left( z \right) = \exp \left( { - \frac{1}{2}\int\limits_{ - \infty }^z {W\left( {z'} \right){\rm d}z'} } \right) .
\end{equation}
Notice that $w\left(\rho\right)$ in \eqref{singular_potential} can be represented as
\begin{equation}
    w\left( \rho  \right) = \frac{{\rm d}}{{{\rm d}\rho }}\left[ {2\ln \left( {\sqrt {1 - \frac{1}{\rho }} } \right) - \ln \left( {\varepsilon  + \mu \sqrt {1 - \frac{1}{\rho }} } \right)} \right].
\end{equation}
Using the last relation, one  obtains
\begin{equation}
\begin{split}
   &- \frac{1}{2}\int\limits_{ - \infty }^z {W\left( {z'} \right){\rm d}z'}  = \\
     & = - \frac{1}{2}\left. {\left[ {\ln \left( {1 - \frac{1}{{\rho \left( {z'} \right)}}} \right) - \ln \left( {\varepsilon  + \mu \sqrt {1 - \frac{1}{{\rho \left( {z'} \right)}}} } \right) - \ln \left( {\frac{{{\rm d}\rho \left( {z'} \right)}}{{{\rm d}z'}}} \right)} \right]} \right|_{ - \infty }^z  =  \\
   &=  - \ln \left( {\sqrt {\frac{{\left( {\rho \left( z \right) - 1} \right)\,\varepsilon }}{{\rho \left( z \right){\textstyle{{{\rm d}\rho \left( z \right)} \over {{\rm d}z}}}\left( {\varepsilon  + \mu \sqrt {1 - {\textstyle{1 \over {\rho \left( z \right)}}}} } \right)}}} } \right).
\end{split}
\end{equation}
Thus, we find
\begin{equation}
    A\left( z \right) = \sqrt {1 + \frac{\mu }{\varepsilon }\sqrt {1 - \frac{1}{{\rho \left( z \right)}}} } .
\end{equation}
Substituting (\ref{f_Au}) into equation (\ref{2_order_f_z}), we finally get
\begin{equation}\label{smooth_eq}
    - \frac{{{\rm d}^{\rm 2} u}}{{{\rm d}z^2 }} + V_\kappa ^{(u)} \left( {\varepsilon ,z} \right)\,u = \varepsilon ^2 u,
\end{equation}
where
\begin{equation} \label{V_u}
\begin{split}
 &V_\kappa ^{(u)} \left( {\varepsilon ,z} \right) = \frac{1}{2}\frac{{{\rm d}W\left( z \right)}}{{{\rm d}z}} + \frac{{W^2 \left( z \right)}}{4} - C\left( z \right) + \varepsilon ^2  =  \\
 &= \frac{{\mu \left( {\rho \left( z \right) - 1} \right)^{{\textstyle{3 \over 2}}}  - \kappa \varepsilon \,\rho \left( z \right)\sqrt {\rho \left( z \right) - 1} }}{{2\rho ^{{\textstyle{9 \over 2}}} \left( z \right)\left( {\varepsilon  + \mu \sqrt {{\textstyle{{\rho \left( z \right) - 1} \over {\rho \left( z \right)}}}} } \right)}} + \frac{{\mu ^2 \left( {\rho \left( z \right) - 1} \right) - 2\mu \varepsilon \sqrt {\left( {\rho \left( z \right) - 1} \right)\rho \left( z \right)} }}{{16\rho ^5 \left( z \right)\left( {\varepsilon  + \mu \sqrt {{\textstyle{{\rho \left( z \right) - 1} \over {\rho \left( z \right)}}}} } \right)^2 }} \, +  \\
 &\hphantom{= \ }
 + \mu ^2 \frac{{\rho \left( z \right) - 1}}{{\rho \left( z \right)}} + \kappa \frac{{\left( {\rho \left( z \right) - 1} \right)^{{\textstyle{3 \over 2}}} }}{{\rho ^{{\textstyle{7 \over 2}}} \left( z \right)}} + \kappa ^2 \frac{{\rho \left( z \right) - 1}}{{\rho ^3 \left( z \right)}}.
\end{split}
\end{equation}
Using formula \eqref{V_u}, one can check that for
any $\varepsilon>0$, $\kappa$ and $\mu$ the energy-dependent
quasipotential $V_\kappa ^{(u)} \left( {\varepsilon ,z} \right)$
is smooth and  behaves at  $z\to\pm\infty$ as follows:
$$\mathop {\lim }\limits_{z \to - \infty } V_\kappa ^{(u)} \left(
{\varepsilon ,z} \right) = 0,\qquad \mathop {\lim }\limits_{z \to
\infty } V_\kappa ^{(u)} \left( {\varepsilon ,z} \right) = \mu ^2.
$$ Having a solution for the function $f_{jl} \left( {\varepsilon
,z} \right)$, one can obtain the solution for the function
$g_{jl'} \left( {\varepsilon ,z} \right)$ using equation
(\ref{dim_less_f}).

\begin{figure}[!ht]
\begin{minipage}[h]{0.49\linewidth}
\begin{center}
\includegraphics[width=1\linewidth]{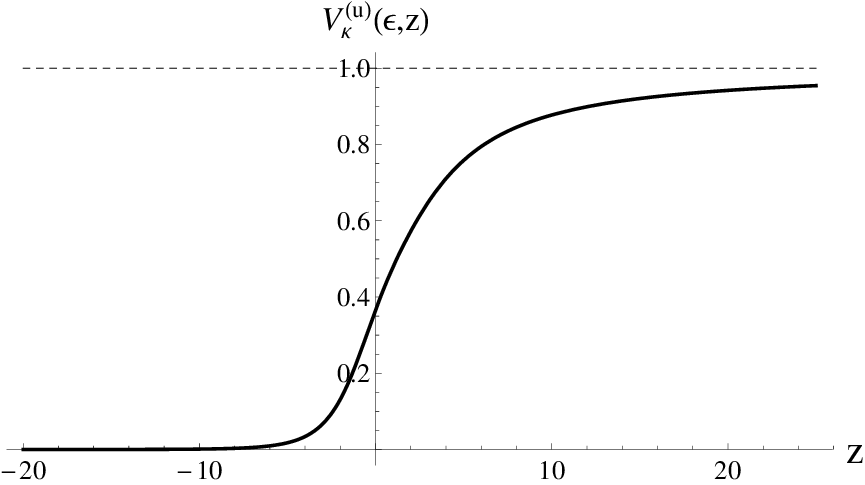}
\end{center}
\end{minipage}
\hfill
\begin{minipage}[h]{0.49\linewidth}
\begin{center}
\includegraphics[width=1\linewidth]{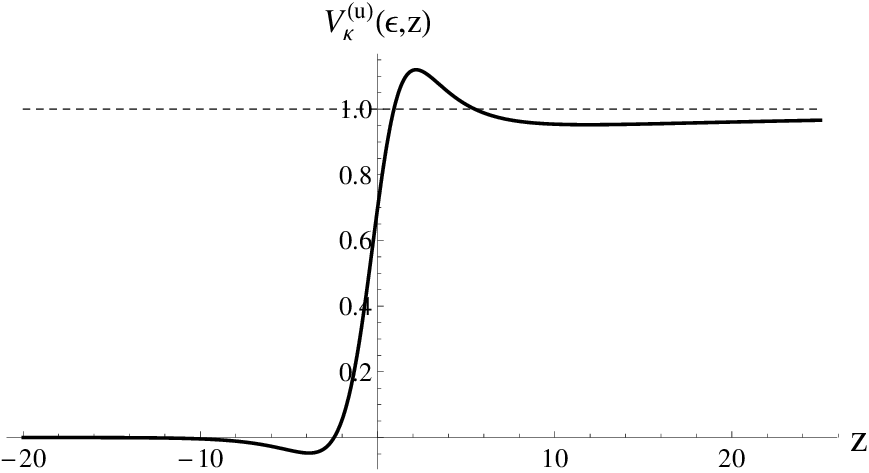}
\end{center}
\end{minipage}
\caption{$V_\kappa ^{(u)} \left( {\varepsilon ,z} \right)$ for $\mu=1$, $\kappa=-1$, $\varepsilon =0.5$ (left plot) and $\mu=1$, $\kappa=2$, $\varepsilon=2$ (right plot). The dotted line corresponds to $\mathop {\lim }\limits_{z \to \infty } V_\kappa ^{(u)} \left( {\varepsilon ,z} \right) = \mu ^2$.}
\label{Vs}
\end{figure}

\begin{figure}[!ht]
\begin{minipage}[h]{0.49\linewidth}
\begin{center}
\includegraphics[width=1\linewidth]{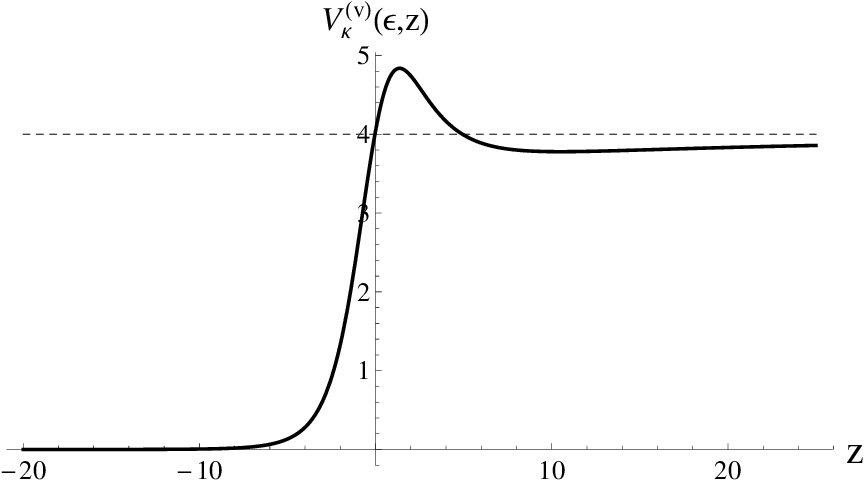}
\end{center}
\end{minipage}
\hfill
\begin{minipage}[h]{0.49\linewidth}
\begin{center}
\includegraphics[width=1\linewidth]{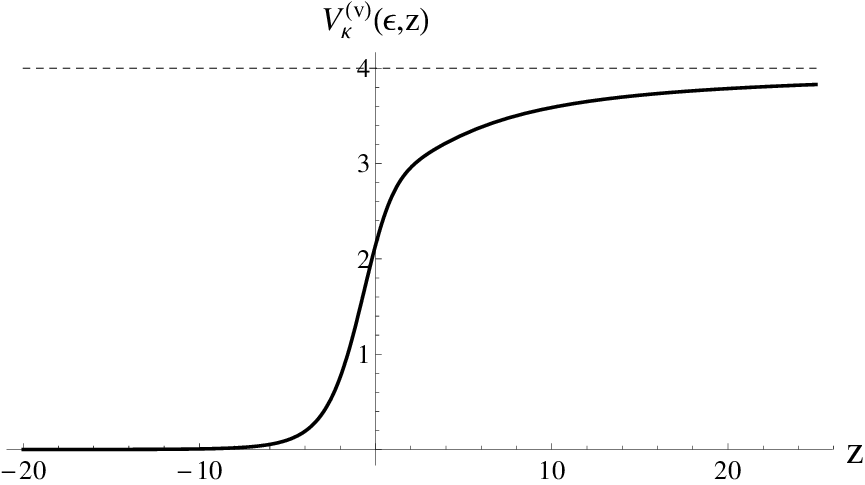}
\end{center}
\end{minipage}
\caption{$V_\kappa ^{(v)} \left( {\varepsilon ,z} \right)$ for $\mu=2$, $\kappa=5$, $\varepsilon=-0.2$ (left plot) and $\mu=2$, $\kappa=3$, $\varepsilon=-3$ (right plot). The dotted line corresponds to $\mathop {\lim }\limits_{z \to \infty } V_\kappa ^{(v)} \left( {\varepsilon ,z} \right) = \mu ^2$.}
\label{Vs2}
\end{figure}

Now let us consider the case $\varepsilon  < 0$. It  should be
noted that equations (\ref{dim_less_f}) and (\ref{dim_less_g}) are
symmetric with respect to the change $f \leftrightarrow g$,
$\varepsilon  \to  - \varepsilon $, $\kappa  \to  - \kappa $. Then
for the function $v = \left(A\left( z
\right)\right)^{-1}\,g,$ defined similar to the function $u$ in
(\ref{f_Au}), we obtain
\begin{equation}\label{smooth_eq_v}
    - \frac{{{\rm d}^{\rm 2} v}}{{{\rm d}z^2 }} + V_\kappa ^{(v)} \left( {\varepsilon ,z} \right)\,v = \varepsilon ^2 v,
\end{equation}
where
\begin{equation} \label{V_v}
    V_\kappa ^{(v)} \left( {\varepsilon ,z} \right) = V_{ - \kappa }^{(u)} \left( { - \varepsilon ,z} \right)
\end{equation}
(it is necessary to work with the function $g_{jl'} \left(
{\varepsilon ,z} \right)$,  since for some
values of $\varepsilon  < 0$ the expression $\varepsilon  + \mu
\sqrt {{\textstyle{{\rho \left( z \right) - 1} \over {\rho \left(
z \right)}}}} $ vanishes at a certain value of $\rho \left( z
\right)$ and leads to the singularity in (\ref{V_u}), while
$\varepsilon  - \mu \sqrt {{\textstyle{{\rho \left( z \right) - 1}
\over {\rho \left( z \right)}}}}$ does not vanish).
Because of \eqref{V_v}, for any $\varepsilon<0$,
$\kappa$ and $\mu$ the quasipotential $V_\kappa ^{(v)} \left(
{\varepsilon ,z} \right)$ has the same asymptotic behavior as
$V_\kappa ^{(u)} \left( {\varepsilon ,z} \right)$:
$$\mathop
{\lim }\limits_{z \to  - \infty } V_\kappa ^{(v)} \left(
{\varepsilon ,z} \right) = 0,\qquad \mathop {\lim }\limits_{z \to
\infty } V_\kappa ^{(v)} \left( {\varepsilon ,z} \right) = \mu ^2.
$$
Having a solution for the function $g_{jl'} \left( {\varepsilon
,z} \right)$, one can obtain the solution for the function $f_{jl}
\left( {\varepsilon ,z} \right)$ using equation
(\ref{dim_less_g}). Examples of quasipotentials (\ref{V_u}),
(\ref{V_v}) are shown in figures~\ref{Vs},~\ref{Vs2}.


Thus, we see that the quasipotentials are a kind
of the one-dimensional  step potential in non-relativistic quantum
mechanics. Note that, contrary to the ordinary quantum mechanics,
in which one has one and the same potential for any eigenvalue,
here for any ``eigenvalue'' $\varepsilon^{2}$ in (\ref{smooth_eq})
(for $\varepsilon>0$) there exists the unique smooth potential
$V_\kappa^{(u)}\left({\varepsilon,z}\right)$. Analogously, for any
``eigenvalue'' $\varepsilon^{2}$ in (\ref{smooth_eq_v}) (for
$\varepsilon<0$) there exists the unique smooth potential
$V_\kappa^{(v)}\left({\varepsilon,z}\right)$. According to the
asymptotic behavior of the quasipotentials, equations
(\ref{smooth_eq}) and (\ref{smooth_eq_v}) have the same continuous
spectra $0<\varepsilon^{2}<\infty$ as the standard equation with
the step potential. As shown in figures~\ref{Vs},~\ref{Vs2},
quasipotentials can look different for differen values of the
parameters; in particular, there can be rises and dips in the
vicinity of $z=0$ which are  consequences of the complicated
definition of the quasipotential \eqref{V_u} and seem not to have
obvious physical interpretations in general. For example, on the
right plot in figure~\ref{Vs} one can see a local minimum, which
looks like a potential well. Note that there cannot be a bound
state in such a potential well, because the corresponding
quasipotentials can be used only for the ``eigenvalues''
$\varepsilon^{2}>0$ (in particular, the quasipotential on the
right plot in figure~\ref{Vs} can be used only for
$\varepsilon^{2}=4$, whereas an assumed bound state in the
potential well implies $\varepsilon^{2}<0$, which is impossible).
Since equations (\ref{smooth_eq}) and (\ref{smooth_eq_v}) are
second order differential equations, one expects that formally
there exist two linearly independent solutions of each equation.
By analogy with the case of the scalar field
\cite{Egorov:2022hgg}, the step-like form of the quasipotentials
suggests that physically relevant solutions are supposed to be
bounded for $z\to\pm\infty$. Thus, for $|\varepsilon|>\mu$, we are
left with two solutions, which oscillate for $z\to\pm\infty$. For
$|\varepsilon|<\mu$, we are left with only one solution, which
falls off exponentially for $z\to\infty$ (the second linearly
independent solution grows exponentially for $z\to\infty$ and
should be discarded as unphysical).

The transition to the functions $u$ and $v$ is needed only in
order to find the energy spectrum of the equations from the form
of the quasipotentials, but it is more convenient to work with the
functions $f$ and $g$. As was demonstrated above, the shape of the quasipotentials
suggests that for $|\varepsilon | > \mu $  there is a pair of
orthogonal solutions $f_{jl}^{(1)} \left( {\varepsilon ,z}
\right)$, $g_{jl'}^{(1)} \left( {\varepsilon ,z} \right)$ and
$f_{jl}^{(2)} \left( {\varepsilon ,z} \right)$, $g_{jl'}^{(2)}
\left( {\varepsilon ,z} \right)$ for each value of $\varepsilon $,
and for $|\varepsilon | < \mu $ there is only one solution $f_{jl}
\left( {\varepsilon ,z} \right)$, $g_{jl'} \left( {\varepsilon ,z}
\right)$ for each value of $\varepsilon $. Then the orthogonality
conditions have the form
\begin{equation}
    \int\limits_{ - \infty }^\infty  {\left( {f_{jl} \left( {\varepsilon ,z} \right),\,\,\,g_{jl'} \left( {\varepsilon ,z} \right)} \right) \cdot \left( {\begin{array}{*{20}c}
   {f_{jl}^{(p)} \left( {\varepsilon ',z} \right)}  \\
   {g_{jl'}^{(p)} \left( {\varepsilon ',z} \right)}  \\
\end{array}} \right){\rm d}z}  = 0,
\end{equation}
\begin{equation}
    \int\limits_{ - \infty }^\infty  {\left( {f_{jl}^{(p)} \left( {\varepsilon ,z} \right),\,\,\,g_{jl'}^{(p)} \left( {\varepsilon ,z} \right)} \right) \cdot \left( {\begin{array}{*{20}c}
   {f_{jl}^{(p')} \left( {\varepsilon ',z} \right)}  \\
   {g_{jl'}^{(p')} \left( {\varepsilon ',z} \right)}  \\
\end{array}} \right){\rm d}z}  = \delta _{pp'} \,\delta \left( {\varepsilon  - \varepsilon '} \right).
\end{equation}

It should be noted here that in paper \cite{Neznamov:2018zen}
equation (\ref{2_order_f_rho}) in the Schwarzschild coordinates
was brought to the Schr\"odinger form with a quasipotential
containing terms proportional to $1/{\left( {\rho - 1} \right)^2}
$, one of such terms being already present in
(\ref{singular_potential}). In non-relativistic quantum mechanics,
such singular potentials lead to the so-called ``fall''\, of a
particle to the center, i.e., to the existence of energy levels
with $\varepsilon \to - \infty$ (see\cite{LL-QM}, Chapter V).
However, by passing to the tortoise coordinate in
this equation, the horizon is moved to $z\to-\infty$ and we
obtain equation (\ref{smooth_eq}) with a smooth quasipotential
for which the energy levels are non-negative. It means that in the
relativistic theory under consideration the singularity of the
quasipotential  at $\rho\to 1$ in  Schwarzschild coordinates is an
artifact of using metric (\ref{metric_Sch}) and
does not indicate any physical pathology of the theory.

Now let us restore the dimensional variables and return to the
original Schwarzschild coordinates. The condition of completeness
of the eigenfunction system of the operator $H_r \left( \kappa
\right)$ can be written as follows:
\begin{equation}
\begin{split}
  &\int\limits_0^m {{\rm d}E\left[ {\left( {\begin{array}{*{20}c}
   {f_{jl} \left( {E,r} \right)}  \\
   {g_{jl'} \left( {E,r} \right)}  \\
\end{array}} \right) \cdot \left( {\begin{array}{*{20}c}
   {f_{jl} \left( {E,r'} \right)}  \\
   {g_{jl'} \left( {E,r'} \right)}  \\
\end{array}} \right)^{\rm T}   + \left( {\begin{array}{*{20}c}
   {g_{jl} \left( {E,r} \right)}  \\
   {f_{jl'} \left( {E,r} \right)}  \\
\end{array}} \right) \cdot \left( {\begin{array}{*{20}c}
   {g_{jl} \left( {E,r'} \right)}  \\
   {f_{jl'} \left( {E,r'} \right)}  \\
\end{array}} \right)^{\rm T}  } \right]}  +  \\
  &+ \sum\limits_{p = 1}^2 {\int\limits_m^\infty  {{\rm d}E\left[ {\left( {\begin{array}{*{20}c}
   {f_{jl}^{(p)} \left( {E,r} \right)}  \\
   {g_{jl'}^{(p)} \left( {E,r} \right)}  \\
\end{array}} \right) \cdot \left( {\begin{array}{*{20}c}
   {f_{jl}^{(p)} \left( {E,r'} \right)}  \\
   {g_{jl'}^{(p)} \left( {E,r'} \right)}  \\
\end{array}} \right)^{\rm T} } \right.}} + \\
  &\hphantom{+ \sum\limits_{p = 1}^2 {\int\limits_m^\infty  {{\rm d}E\Bigg[ \ \, }}}
    \left. { + \left( {\begin{array}{*{20}c}
   {g_{jl}^{(p)} \left( {E,r} \right)}  \\
   {f_{jl'}^{(p)} \left( {E,r} \right)}  \\
\end{array}} \right) \cdot \left( {\begin{array}{*{20}c}
   {g_{jl}^{(p)} \left( {E,r'} \right)}  \\
   {f_{jl'}^{(p)} \left( {E,r'} \right)}  \\
\end{array}} \right)^{\rm T}  } \right]   =  \\
  &= \left( {\begin{array}{*{20}c}
   1 & 0  \\
   0 & 1  \\
\end{array}} \right)\left( {1 - \frac{{r_0 }}{r}} \right)\delta \left( {r - r'} \right).
\end{split}
\end{equation}
In components it reads:
\begin{equation}
\begin{split}
  &\int\limits_0^m {{\rm d}E\left( {f_{jl} \left( {E,r} \right)f_{jl} \left( {E,r'} \right) + g_{jl} \left( {E,r} \right)g_{jl} \left( {E,r'} \right)} \right)} \, +  \\
  &+ \sum\limits_{p = 1}^2 {\int\limits_m^\infty  {{\rm d}E\left( {f_{jl}^{(p)} \left( {E,r} \right)f_{jl}^{(p)} \left( {E,r'} \right) + g_{jl}^{(p)} \left( {E,r} \right)g_{jl}^{(p)} \left( {E,r'} \right)} \right)} }  = \left( {1 - \frac{{r_0 }}{r}} \right)\delta \left( {r - r'} \right),
\end{split}
\end{equation}
\begin{equation}
\begin{split}
  &\int\limits_0^m {{\rm d}E\left( {f_{jl} \left( {E,r} \right)g_{jl'} \left( {E,r'} \right) + g_{jl} \left( {E,r} \right)f_{jl'} \left( {E,r'} \right)} \right)} \, +  \\
  &+ \sum\limits_{p = 1}^2 {\int\limits_m^\infty  {{\rm d}E\left( {f_{jl}^{(p)} \left( {E,r} \right)g_{jl'}^{(p)} \left( {E,r'} \right) + g_{jl}^{(p)} \left( {E,r} \right)f_{jl'}^{(p)} \left( {E,r'} \right)} \right)} }  = 0.
\end{split}
\end{equation}
Returning to the functions $F_{jl}$ and $G_{jl}$, we obtain the completeness condition for them in the form
\begin{equation} \label{completeness_FG}
\begin{split}
  &\int\limits_0^m {{\rm d}E\left( {F_{jl} \left( {E,r} \right)F_{jl} \left( {E,r'} \right) + G_{jl} \left( {E,r} \right)G_{jl} \left( {E,r'} \right)} \right)} \, +  \\
  &+ \sum\limits_{p = 1}^2 {\int\limits_m^\infty  {{\rm d}E\left( {F_{jl}^{(p)} \left( {E,r} \right)F_{jl}^{(p)} \left( {E,r'} \right) + G_{jl}^{(p)} \left( {E,r} \right)G_{jl}^{(p)} \left( {E,r'} \right)} \right)} }  = \frac{{\sqrt {1 - {\textstyle{{r_0 } \over r}}} }}{{r^2 }}\delta \left( {r - r'} \right),
\end{split}
\end{equation}
\begin{equation} \label{completeness_FG_0}
\begin{split}
  &\int\limits_0^m {{\rm d}E\left( {F_{jl} \left( {E,r} \right)G_{jl'} \left( {E,r'} \right) + G_{jl} \left( {E,r} \right)F_{jl'} \left( {E,r'} \right)} \right)} \, +  \\
  &+ \sum\limits_{p = 1}^2 {\int\limits_m^\infty  {{\rm d}E\left( {F_{jl}^{(p)} \left( {E,r} \right)G_{jl'}^{(p)} \left( {E,r'} \right) + G_{jl}^{(p)} \left( {E,r} \right)F_{jl'}^{(p)} \left( {E,r'} \right)} \right)} }  = 0.
\end{split}
\end{equation}

In the next section we will use these conditions for proving the
canonical commutation relation for the spinor field.

Above we have shown that the Dirac Hamiltonian
corresponding to equation (\ref{curved_Dirac}) has only the continuous
spectrum stretching from $-\infty$ to $\infty$ and its
eigenfunctions can be  built from the functions $F_{jl}$, $G_{jl}$
and the spherical spinors as follows. The solutions corresponding
to the positive eigenvalues with energy $E < m$ will be denoted by
$ \phi _{Ejlm_{\rm t} } \left( {r,\theta ,\varphi } \right)$ and
are given by
\begin{equation}\label{sol_pos_fin}
    \phi _{Ejlm_{\rm t} } \left( {r,\theta ,\varphi } \right) = \left( {\begin{array}{*{20}c}
   {F_{jl} \left( {E,r} \right)\Omega _{jlm_{\rm t} } \left( {\theta ,\varphi } \right)}  \\
   {{\rm i}G_{jl'} \left( {E,r} \right)\Omega _{jl'm_{\rm t} } \left( {\theta ,\varphi } \right)}  \\
\end{array}} \right).
\end{equation}

As it has been noted above, due to the symmetry of
equations (\ref{radial_F}) and (\ref{radial_G})  with respect to the
permutation $F \leftrightarrow G$, $E \to  - E$, $\kappa  \to -
\kappa $ (or $l \leftrightarrow l'$), we can immediately represent
the negative energy solution with $|E| < m$, which will be denoted
by $\chi _{Ejlm_{\rm t} } \left( {r,\theta ,\varphi } \right)$, in
the form
\begin{equation}\label{sol_neg_fin}
    \chi _{Ejlm_{\rm t} } \left( {r,\theta ,\varphi } \right) = \left( {\begin{array}{*{20}c}
   {G_{jl} \left( {E,r} \right)\Omega _{jlm_{\rm t} } \left( {\theta ,\varphi } \right)}  \\
   {{\rm i}F_{jl'} \left( {E,r} \right)\Omega _{jl'm_{\rm t} } \left( {\theta ,\varphi } \right)}  \\
\end{array}} \right).
\end{equation}

The doubly degenerate solutions with positive and negative
energy  $|E| > m$ will be denoted by  $ \phi
_{Ejlm_{\rm t} }^{(p)} \left( {r,\theta ,\varphi } \right)$ and $
\chi _{Ejlm_{\rm t} }^{(p)} \left( {r,\theta ,\varphi } \right)$,
$p = 1,2$, respectively, and are expressed in terms of the
functions $F_{jl}^{(p)}$ and $G_{jl}^{(p)}$  in the same way, as
the functions $ \phi _{Ejlm_{\rm t} } \left( {r,\theta ,\varphi }
\right)$ and $ \chi _{Ejlm_{\rm t} } \left( {r,\theta ,\varphi }
\right)$ in formulas (\ref{sol_pos_fin}), (\ref{sol_neg_fin}).

The normalization conditions for these solutions can be easily
obtained from normalization conditions (\ref{radial_F}) and the
normalization conditions for the spherical spinors
(\ref{norm_Omega}). The former imply that all these solutions have
infinite norm. We also remind that this system of
solutions is complete due to the Dirac Hamiltonian being
Hermitian.

\section{Canonical quantization}
Now we can perform the canonical quantization of
the spinor field.  It is most convenient to carry out the
quantization procedure using the functions $f$ and $g$ and
performing calculations in the tortoise coordinate (due to the
simple orthogonality and completeness conditions). Nevertheless,
in what follows we will give formulas in the more familiar
Schwarzschild coordinates.

The expansion of the spinor field in terms of the complete system
of stationary solutions of Dirac equation (\ref{curved_Dirac})
constructed in the previous section has the form
\begin{equation} \label{decomp}
\begin{split}
  &\psi \left( {t,r,\theta ,\varphi } \right) = \sum\limits_{j = {\textstyle{1 \over 2}}}^\infty  {\sum\limits_{m_{\rm t}  =  - j}^j {\sum\limits_{l = j \pm {\textstyle{1 \over 2}}} } } \\
  &\left[ { {\int\limits_0^m {{\rm d}E\left( {{\rm e}^{ - {\rm i}Et} \phi _{Ejlm_{\rm t} } \left( {r,\theta ,\varphi } \right)a_{jlm_{\rm t} } \left( E \right) + {\rm e}^{{\rm i}Et} \chi _{Ejlm_{\rm t} } \left( {r,\theta ,\varphi } \right)b_{jlm_{\rm t} }^ \dagger  \left( E \right)} \right)} }  + } \right. \\
  &\ \ \left. {+ \sum\limits_{p = 1}^2  {\int\limits_m^\infty  {{\rm d}E\left( {{\rm e}^{ - {\rm i}Et} \phi _{Ejlm_{\rm t} }^{(p)} \left( {r,\theta ,\varphi } \right)a_{jlm_{\rm t} }^{(p)} \left( E \right) + {\rm e}^{{\rm i}Et} \chi _{Ejlm_{\rm t} }^{(p)} \left( {r,\theta ,\varphi } \right)b_{jlm_{\rm t} }^{(p) \dagger } \left( E \right)} \right)} } } \right].
\end{split}
\end{equation}

The creation and annihilation operators $a$, $b$, $a^{(p)}$, $b^{(p)}$ must satisfy the anticommutation rules
\begin{equation} \label{anticommutators}
\begin{split}
  \left\{ {a_{jlm_{\rm t} } \left( E \right),a_{j'l'm'_{\rm t} }^ \dagger  \left( {E'} \right)} \right\} &= \delta _{jj'} \,\delta _{ll'} \,\delta _{m_{\rm t} m'_{\rm t} } \,\delta \left( {E - E'} \right), \\
  \left\{ {b_{jlm_{\rm t} } \left( E \right),b_{j'l'm'_{\rm t} }^ \dagger  \left( {E'} \right)} \right\} &= \delta _{jj'} \,\delta _{ll'} \,\delta _{m_{\rm t} m'_{\rm t} } \,\delta \left( {E - E'} \right), \\
  \left\{ {a_{jlm_{\rm t} }^{(p)} \left( E \right),a_{j'l'm'_{\rm t} }^{(p') \dagger } \left( {E'} \right)} \right\} &= \delta _{pp'} \,\delta _{jj'} \,\delta _{ll'} \,\delta _{m_{\rm t} m'_{\rm t} } \,\delta \left( {E - E'} \right), \\
  \left\{ {b_{jlm_{\rm t} }^{(p)} \left( E \right),b_{j'l'm'_{\rm t} }^{(p') \dagger } \left( {E'} \right)} \right\} &= \delta _{pp'} \,\delta _{jj'} \,\delta _{ll'} \,\delta _{m_{\rm t} m'_{\rm t} } \,\delta \left( {E - E'} \right).
\end{split}
\end{equation}

Now, using these anticommutation relations, the properties of the
radial functions and spherical spinors, we obtain the
anticommutation conditions for the spinor fields $\psi \left(
{t,r,\theta ,\varphi } \right)$ and $\psi ^ \dagger  \left(
{t,r',\theta ',\varphi '} \right)$. Using expansion (\ref{decomp})
and relations (\ref{anticommutators}), and then completeness
condition (\ref{completeness_FG}) for the radial functions and
completeness condition (\ref{completeness_Omega}) for the
spherical spinors, one gets
\begin{equation} \label{canon_anticom}
    \left\{ {\psi _\alpha  \left( {t,r,\theta ,\varphi } \right),\psi _\beta ^ \dagger  \left( {t,r',\theta ',\varphi '} \right)} \right\} = \delta _{\alpha \beta } \frac{{\sqrt {1 - {\textstyle{{r_0 } \over r}}} }}{{r^2 }}\delta \left( {r - r'} \right)\delta \left( {\cos \theta  - \cos \theta '} \right)\delta \left( {\varphi  - \varphi '} \right)
\end{equation}
($\alpha ,\beta  = 1,2,3,4$ are the spinor indices). Thus, we
arrive at the canonical anticommutation relations in curved
spacetime.

We emphasize once again that since there is no translation
invariance in the Schwarzschild spacetime, wave functions of
fermions do not decompose into the product of a
coordinate-dependent exponential function and a spinor that is a
function of only the momentum. In the basis of the eigenfunctions
of the squared angular momentum operator $\vec J^2$, which we must
use, (anti)fermion states are described by coordinate-dependent
spinors. Therefore, the orthogonality (\ref{norm_FG}) and
completeness (\ref{completeness_FG}) and (\ref{completeness_FG_0})
conditions of the radial functions and completeness condition
(\ref{completeness_Omega}) of the system of spherical spinors are
crucial for the derivation of canonical anticommutation relations
(\ref{canon_anticom}) and play the role of the spin sums in
Minkowski space.

Finally, let us find the Hamiltonian of the spinor field  in the
gravitational field of a Schwarzschild black hole. The Hamiltonian
density corresponding to the Lagrangian $L$ defined in
(\ref{fermion_action}) is determined by the formula
\begin{equation}
    \tilde H = \frac{{\partial L}}{{\partial \dot \psi }}\dot \psi  + \dot {\bar \psi} \frac{{\partial L}}{{\partial \dot {\bar \psi} }} - L.
\end{equation}
Using the diagonal property of the tetrad $e_{(\nu )}^\mu$, we can represent the derivatives in the form
\begin{equation}
    \frac{{\partial L}}{{\partial \dot \psi }} = \frac{{\rm i}}{2}\sqrt { - g} \bar \psi \gamma ^{(0)} e_{(0)}^0 , \qquad \frac{{\partial L}}{{\partial \dot {\bar \psi} }} =  - \frac{{\rm i}}{2}\sqrt { - g} \, e_{(0)}^0 \gamma ^{(0)} \psi .
\end{equation}
Taking  into account that the fermionic Lagrangian is equal to
zero on the equations of motion,  $\left. L \right|_{{\rm EoM}}  =
0$, we obtain  the Hamiltonian density in the form
\begin{equation} \label{Hamiltonian_density}
    \tilde H = \frac{{\rm i}}{2}\sqrt { - g} \, e_{(0)}^0 \left( {\bar \psi \gamma ^{(0)} \dot \psi  - \dot {\bar \psi} \gamma ^{(0)} \psi } \right) = \frac{{\rm i}}{2}\sqrt { - g} \, e_{(0)}^0 \left( {\psi ^ \dagger  \dot \psi  - \dot \psi ^ \dagger  \psi } \right).
\end{equation}
Expression (\ref{Hamiltonian_density}) agrees with the expression
for  the component $T_{00}$ of the energy-momentum tensor of the
spinor field in curved spacetime given in monograph~\cite{BD}.
Taking into account that for the Schwarzschild
metric
\begin{equation}
    \sqrt { - g}  = r^2 \sin \theta ,
\end{equation}
the tetrad in the Schwinger gauge is given by
\begin{equation}
    e_{(0)}^0  = \frac{1}{{\sqrt {1 - {\textstyle{{r_0 } \over r}}} }},
\end{equation}
and passing from the Hamiltonian density $\tilde H$ to the
Hamiltonian $H$ proper, one obtains
\begin{equation}
    H = \int {\tilde H\,{\rm d}^3 x}  = \frac{{\rm i}}{2}\int {\frac{{r^2 \sin \theta }}{{\sqrt {1 - {\textstyle{{r_0 } \over r}}} }}\left( {\psi ^ \dagger  \dot \psi  - \dot \psi ^ \dagger  \psi } \right){\rm d}r\,{\rm d}\theta \,{\rm d}\varphi } .
\end{equation}

Let us now substitute the expansion of the spinor field $\psi$ in
form (\ref{decomp})  into this formula. Applying orthogonality
condition (\ref{norm_Omega}) for the spherical spinors and then
orthogonality condition (\ref{norm_FG}) for the radial functions
$F$ and $G$, as a result of not difficult but bulky calculations,
we arrive at the following expression for the Hamiltonian:
\begin{equation}
\begin{split}
 H = \frac{1}{2}\sum\limits_{j = {\textstyle{1 \over 2}}}^\infty  {\sum\limits_{m_{\rm t}  =  - j}^j {\sum\limits_{l = j \pm {\textstyle{1 \over 2}}} {\left( {\int\limits_0^m {E\left( {a_{jlm_{\rm t} }^ \dagger  \left( E \right)a_{jlm_{\rm t} } \left( E \right) - b_{jlm_{\rm t} } \left( E \right)b_{jlm_{\rm t} }^ \dagger  \left( E \right)} \right){\rm d}E} \, + } \right.} } }  \\
 \left. { + \sum\limits_{p = 1}^2 {\int\limits_m^\infty  {E\left( {a_{jlm_{\rm t} }^{(p) \dagger } \left( E \right)a_{jlm_{\rm t} }^{(p)} \left( E \right) - b_{jlm_{\rm t} }^{(p)} \left( E \right)b_{jlm_{\rm t} }^{(p) \dagger } \left( E \right)} \right){\rm d}E} } } \right) -\\
 -\frac{1}{2}\sum\limits_{j = {\textstyle{1 \over 2}}}^\infty  {\sum\limits_{m_{\rm t}  =  - j}^j {\sum\limits_{l = j \pm {\textstyle{1 \over 2}}} {\left( {\int\limits_0^m {E\left(a_{jlm_{\rm t} } \left( E \right){a_{jlm_{\rm t} }^ \dagger  \left( E \right) -b_{jlm_{\rm t} }^ \dagger  \left( E \right)b_{jlm_{\rm t} } \left( E \right)} \right){\rm d}E} \, + } \right.} } }  \\
 \left. { + \sum\limits_{p = 1}^2 {\int\limits_m^\infty  {E\left( {a_{jlm_{\rm t} }^{(p)} \left( E \right)a_{jlm_{\rm t} }^{(p) \dagger } \left( E \right) - b_{jlm_{\rm t} }^{(p) \dagger } \left( E \right)b_{jlm_{\rm t} }^{(p)} \left( E \right)} \right){\rm d}E} } } \right).
 \end{split}
\end{equation}
Passing here to the normal ordering of the creation and
annihilation  operators and dropping infinite c-number terms, one
obtains the Hamiltonian of the system in the final form:
\begin{equation}
\begin{split}
 H = \sum\limits_{j = {\textstyle{1 \over 2}}}^\infty  {\sum\limits_{m_{\rm t}  =  - j}^j {\sum\limits_{l = j \pm {\textstyle{1 \over 2}}} {\left( {\int\limits_0^m {E\left( {a_{jlm_{\rm t} }^ \dagger  \left( E \right)a_{jlm_{\rm t} } \left( E \right) + b_{jlm_{\rm t} }^ \dagger  \left( E \right)b_{jlm_{\rm t} } \left( E \right)} \right){\rm d}E} \, + } \right.} } }  \\
 \,\,\,\,\,\,\,\,\,\,\,\,\,\,\,\,\,\,\left. { + \sum\limits_{p = 1}^2 {\int\limits_m^\infty  {E\left( {a_{jlm_{\rm t} }^{(p) \dagger } \left( E \right)a_{jlm_{\rm t} }^{(p)} \left( E \right) + b_{jlm_{\rm t} }^{(p) \dagger } \left( E \right)b_{jlm_{\rm t} }^{(p)} \left( E \right)} \right){\rm d}E} } } \right).
\end{split}
\end{equation}
As in the case of the scalar field \cite{Egorov:2022hgg}, it has
the  standard form, except for integrating separately over the
spectra of finite and infinite motion and doubling the number of
states in the latter case.

\section{Conclusion}
Having carried out the procedure of canonical quantization of  a
massive spinor field outside the horizon of an ideal Schwarzschild
black hole, we have shown that a consistent quantum theory of the
spinor field in the gravitational field of the Schwarzschild black
hole can be constructed, in complete analogy with the case of the
scalar field \cite{Egorov:2022hgg}, without taking into account
the influence of the interior of the black hole (i.e.\ the area
below the horizon). It is shown that the resulting theory is
complete and self-consistent, that is, the canonical commutation
relations hold exactly, and the Hamiltonian has the standard form
without any peculiarities.

Although the time coordinate $t$ can be considered proper  time
only when $r \to \infty$, this does not lead to a contradiction
when performing the canonical quantization procedure. In addition,
since the original theory is invariant under time shifts $t$, it
generates the time-conserved Hamiltonian, which is necessary to
obtain a correct quantum field theory. Thus, in the resulting
quantum theory, time $t$ can be considered as global time.

The spectrum of spinor quantum states, like the spectrum of scalar
states, includes two branches. The first one is a continuous
spectrum of states with energies less than the field mass, which
corresponds to particles trapped in the vicinity of the horizon.
The second branch is a continuous spectrum of states with energies
exceeding the field mass and corresponds to the infinite motion of
particles. The states in this part of the spectrum are doubly
degenerate. It is worth noting that there is no such
degeneracy of spinor particle states with definite energy and
total angular momentum in central potentials in Minkowski
spacetime. We believe that it is due to the topological structure
$\mathbb{R}^2 \times S^2$ of the Schwarzschild spacetime, which differs
from the topological structure $\mathbb{R}^4$ of Minkowski spacetime.
Indeed, in the simpler case of  the scalar field
the topological nature of the degeneracy was demonstrated
explicitly in paper \cite{Egorov:2022hgg} by a
direct comparison with solutions in Minkowski space. Thus, we
expect the appearance of analogous degeneracies in other curved
spacetimes like the one of wormholes (see the discussion of
traversable wormholes of the Morris-Thorne type
\cite{Ellis:1973yv,Bronnikov:1973fh,Morris:1988cz} in paper
\cite{Egorov:2022hgg}) or the Kerr spacetime.

Note that the above procedure of the canonical  quantization of
the spinor field is based on the use of solutions to
the equations of motion in the spherical coordinate system.
However, we assume that, as in the case of the scalar field
\cite{Egorov:2022hgg,Smolyakov}, there is a way first to pass to
the scatteringlike states \cite{Egorov:2022hgg} and then to their
linear combinations \cite{Smolyakov}, which are more convenient
from the point of view of describing the theory at large distances
from the black hole. Unlike the case of the scalar field, which is
very similar to ordinary quantum mechanics from the technical
point of view and, accordingly, in which known results obtained in
quantum mechanics in the framework of the study of scattering
processes can be applied, the case of a spinor field seems to be
more complicated. At the same time, based on the available results
on the properties of the spectrum of radial solutions of the
corresponding equations of motion (namely, taking into account the
twofold degeneracy of the solutions for $E^2 > m^2$), we also
expect the appearance of the twofold degeneracy of the states that
at large distances from the black hole look like having the same
asymptotic momentum. This problem calls for a further
investigation.

\section*{Acknowledgements}
The authors are grateful to E.E.~Boos, V.I.~Denisov, Yu.V.~Grats, S.I.~Keizerov, V.P.~Neznamov and E.R.~Rakhmetov for helpful
discussions. This study was conducted within the scientific
program of the National Center for Physics and Mathematics,
section \textnumero 5 ``Particle Physics and Cosmology''. Stage
2023-2025.


\begin{thebibliography}{99}
\bibitem{Boulware:1974dm}
D.G.~Boulware, {\em ``Quantum Field Theory in Schwarzschild and Rindler Spaces''}, Phys. Rev. D \textbf{11} (1975) 1404.

\bibitem{Hartle:1976tp}
J.B.~Hartle and S.W.~Hawking, {\em ``Path Integral Derivation of Black Hole Radiance''}, Phys. Rev. D \textbf{13} (1976) 2188-2203.

\bibitem{Kruskal:1959vx}
M.D.~Kruskal, {\em ``Maximal extension of Schwarzschild metric''}, Phys. Rev. \textbf{119} (1960) 1743-1745.

\bibitem{Szekeres:1960gm}
G.~Szekeres, {\em ``On the singularities of a Riemannian manifold''}, Publ. Math. Debrecen \textbf{7} (1960) 285-301.

\bibitem{tHooft:2018waj}
G.~'t Hooft, {\em ``Virtual Black Holes and Space\textendash{}Time Structure''}, Found. Phys. \textbf{48} (2018) 1134-1149.

\bibitem{tHooft:2019xwm}
G.~'t Hooft, {\em ``The quantum black hole as a theoretical lab, a pedagogical treatment of a new approach''}, arXiv:1902.10469 [gr-qc].

\bibitem{Hooft:2022azz}
G.~'t.~Hooft, {\em ``Quantum clones inside black holes''}, arXiv:2206.04608 [gr-qc].

\bibitem{Deruelle:1974zy}
N.~Deruelle and R.~Ruffini, {\em ``Quantum and classical relativistic energy states in stationary geometries''},
Phys. Lett. B \textbf{52} (1974) 437-441.

\bibitem{Zecca:2009zz}
A.~Zecca, {\em ``Properties of radial equation of scalar field in Schwarzschild space-time''},
Nuovo Cim. B \textbf{124} (2009) 1251-1258.

\bibitem{Barranco:2011eyw}
J.~Barranco, A.~Bernal, J.~C.~Degollado, A.~Diez-Tejedor, M.~Megevand, M.~Alcubierre, D.~Nunez and O.~Sarbach,
{\em ``Are black holes a serious threat to scalar field dark matter models?''},
Phys. Rev. D \textbf{84} (2011) 083008 [arXiv:1108.0931 [gr-qc]].

\bibitem{Akhmedov:2020ryq}
E.T.~Akhmedov, P.A.~Anempodistov, K.V.~Bazarov, D.V.~Diakonov, U.~Moschella,
{\em ``Heating up an environment around black holes and inside de Sitter space''},
Phys. Rev. D \textbf{103} (2021) 025023
[arXiv:2010.10877 [hep-th]].

\bibitem{Anempodistov:2020oki}
P.A.~Anempodistov, {\em ``Remarks on the thermofield double state in 4D black hole background''},
Phys. Rev. D \textbf{103} (2021) 105008 [arXiv:2012.03305 [hep-th]].

\bibitem{Bazarov:2021rrb}
K.V.~Bazarov, {\em ``Notes on peculiarities of quantum fields in space-times with horizons''},
Class. Quant. Grav. \textbf{39} (2022) 217001 [arXiv:2112.02188 [hep-th]].

\bibitem{Egorov:2022hgg}
V.~Egorov, M.~Smolyakov and I.~Volobuev,
{\em ``Doubling of physical states in the quantum scalar field theory for a remote observer in the Schwarzschild spacetime''},
Phys. Rev. D \textbf{107} (2023) 025001 [arXiv:2209.02067 [gr-qc]].

\bibitem{Zecca:2007}
A.~Zecca, {\em ``Spin 1/2 Bound States in Schwarzschild Geometry''}, Adv. Stud. Theor. Phys. \textbf{1} (2007) 271-279.

\bibitem{Neznamov:2010}
M.V.~Gorbatenko and V.P.~Neznamov, {\em ``Solution of the problem of uniqueness and hermiticity of hamiltonians for Dirac particles in gravitational fields''},
Phys. Rev. D \textbf{82} (2010) 104056 [arXiv:1007.4631 [gr-qc]].

\bibitem{Neznamov:2011}
M.V.~Gorbatenko and V.P.~Neznamov, {\em ``Uniqueness and Self-Conjugacy of Dirac Hamiltonians in arbitrary Gravitational Fields''},
Phys. Rev. D \textbf{83} (2011) 105002 [arXiv:1102.4067 [gr-qc]].

\bibitem{Neznamov:2018zen}
V.P.~Neznamov and I.I.~Safronov, {\em ``Stationary solutions of second-order equations for point fermions in the Schwarzschild gravitational field''},
J. Exp. Theor. Phys. \textbf{127} (2018) 647-658 [arXiv:1809.08940 [gr-qc]].

\bibitem{Neznamov:2020}
M.V.~Gorbatenko and V.P.~Neznamov, {\em ``Quantum mechanics of stationary states of particles in a space\textendash{}time of classical black holes''},
Theor. Math. Phys. \textbf{205} (2020) 1492-1526 [arXiv:2012.04491 [physics.gen-ph]].

\bibitem{Schweber}
S.S.~Schweber, {\em ``An Introduction to Relativistic Quantum Field Theory''}, Dover Publications, 2005.

\bibitem{rigged_Hilbert}
J-P.~Antoine, R.C.~Bishop,  A.~Bohm,  S.~Wickramasekara, {\em ``Rigged
Hilbert Spaces in Quantum Physics''}, In: Greenberger, D.,
Hentschel, K., Weinert, F. (eds) Compendium of Quantum Physics.
Springer, Berlin, Heidelberg, 2009.

\bibitem{Boulware:1975pe}
D.G.~Boulware, {\em ``Spin 1/2 Quantum Field Theory in Schwarzschild Space''}, Phys. Rev. D \textbf{12} (1975) 350.

\bibitem{AB}
A.I.~Akhiezer and V.B.~Berestetskii, {\em ``Quantum electrodynamics''}, Interscience Publishers; Revised Edition, 1965.

\bibitem{LL-QM}
L.D.~Landau and E.M.~Lifshitz, {\em ``Quantum mechanics.
Non-relativistic theory''}, Second edition, Pergamon press, 1965.

\bibitem{BD}
N.D.~Birrell and P.C.W.~Davies, {\em ``Quantum Fields in Curved Space''}, Cambridge Univ. Press, 1984.

\bibitem{Ellis:1973yv}
H.G.~Ellis, {\em ``Ether flow through a drainhole: A particle model in general relativity''}, J. Math. Phys. \textbf{14} (1973) 104.

\bibitem{Bronnikov:1973fh}
K.A.~Bronnikov, {\em ``Scalar-tensor theory and scalar charge''}, Acta Phys. Polon. B \textbf{4} (1973) 251.

\bibitem{Morris:1988cz}
M.S.~Morris, K.S.~Thorne, {\em ``Wormholes in space-time and their use for interstellar travel: A tool for teaching general relativity''},
Am. J. Phys. \textbf{56} (1988) 395.

\bibitem{Smolyakov}
M.N.~Smolyakov, {\em ``Asymptotic behavior of solutions and
spectrum of states in the quantum scalar field theory in the
Schwarzschild spacetime''}, Phys. Rev. D \textbf{108} (2023) 105006 [arXiv:2309.06249 [gr-qc]].
\end{thebibliography}
\end{document}